\documentclass[3p,times,procedia]{elsarticle}
\usepackage{nupha_ecrc}
\usepackage{amssymb}
\usepackage{graphicx}
\usepackage{color}
\usepackage{amsmath,amssymb}

\volume{00}

\firstpage{1}

\journalname{Nuclear Physics A}

\runauth{L.~Yan et al.}

\jid{nupha}

\jnltitlelogo{Nuclear Physics A}

\usepackage{amssymb}

\usepackage[figuresright]{rotating}

\def\be{\begin{eqnarray}}
\def\ee{\end{eqnarray}}

\def\Eq#1{Eq.~(\ref{#1})}
\def\Eqs#1{Eqs.~(\ref{#1})}

\def\Fig#1{Fig.~\ref{#1}}

\def\E{\mathcal{E}}

\def\bra{\langle}
\def\ket{\rangle}
\begin{document}

\begin{frontmatter}



\dochead{}

\title{Nonlinear hydrodynamic response
confronts LHC data}


\author[{a}]{Li Yan}
\author[{b}]{Subrata Pal}
\author[{a}]{and Jean-Yves Ollitrault}
\address[{a}]{Institut de Physique Th\'eorique,
Universit\'e Paris Saclay,
CEA, CNRS, F-91191 Gif-sur-Yvette, France\corref{label1}}
\address[{b}]{Department of Nuclear and Atomic Physics, Tata Institute of
Fundamental Research, Homi Bhabha Road, Mumbai, 400005, India}

\begin{abstract}
Higher order harmonic flow $v_n$ (with $n\ge4$) in heavy-ion collisions 
can be measured either with respect to their own plane, or with respect to a plane 
constructed using lower-order harmonics. 
By assuming that higher flow harmonics are the superposition of medium nonlinear 
and linear responses to initial anisotropies, 
we propose a set of nonlinear response coefficients $\chi_n$'s,
which are independent of initial state by construction.
In experiments, $\chi_n$'s can be extracted as the ratio between 
higher order harmonic flow measured in the plane constructed by $v_2$
and $v_3$, and moments of lower order harmonic flow.
Simulations with single-shot hydrodynamics and AMPT model lead to
results of these nonlinear response coefficients 
in good agreement with the experimental data at the LHC energy.
Predictions for 
$v_7$ and $v_8$ measured with respect to plane of lower order harmonics 
are given accordingly.

\end{abstract}


\end{frontmatter}

\section{Introduction}
\label{sec:intro}

The observed flow phenomena in high energy heavy-ion 
collisions carried out at RHIC and the LHC provides
great opportunities in analyzing the collective dynamics of 
the strongly-coupled Quark-Gluon Plasma (QGP) (for a
recent review, cf.~\cite{Heinz:2013th}). 
To a quantitative level,
the analysis of harmonic flow $V_n$, which
is defined through a Fourier decomposition of the 
observed particle spectrum,
\be
\label{Vndef}
V_n = v_n e^{in\Psi_n} =\{ e^{in\phi_p}\}\,,
\ee
has led to strong constraints on the dissipative properties of 
the QGP medium in various aspects. For instance, 
correlations among flow harmonics have been studied in terms
of the correlations between event-plane $\Psi_n$~\cite{Aad:2014fla}, which
present non-trivial patterns 
depending on the shear viscosity over entropy ratio $\eta/s$ of the medium.
Recent measurements of harmonic flow $V_n$
have achieved results with high precisions, which extends the studies 
of harmonic flow to flow fluctuations~\cite{Aad:2013xma} 
and higher order harmonic flow 
($n\ge4$)~\cite{Chatrchyan:2013kba}. In particular, higher order 
flow harmonics have been measured with respect to their own event-plane,
and plane constructed by lower order flow harmonics, from which nonlinear medium
response to initial eccentricities can be studied. In this work, we focus
on the nonlinear generation of higher order flow harmonics in heavy-ion 
collisions. By assuming that higher harmonics are the superposition of medium nonlinear 
and linear responses, a new set of nonlinear response coefficients
are formulated.

\section{Nonlinear hydrodynamic response and $\chi_n$}
\label{sec:formulation}

We expand harmonic flow 
$V_n$ in a series of initial eccentricities $\E_n$,
accounting for the fact that magnitudes of initial eccentricities are small.
Note that in this work, $\E_n$, as well as $V_n$ defined in \Eq{Vndef} are taken as
complex quantities.
For higher order flow harmonics, it has been shown that nonlinear hydro
response to initial eccentricities result in significant contributions~\cite{Gardim:2011xv,Teaney:2012ke}.  
Taking the fourth order harmonic flow $V_4$ as an example, in addition to the component
$V_4^L$
which is linearly proportional to $\E_4$, there exists a large fraction induced by
hydro response to $\E_2^2$. Therefore, one can write $V_4$ as, 
\be
\label{eq:v4}
V_4 = V_4^L + \chi_4 V_2^2\,.
\ee
$\E_2^2$ has been absorbed into $V_2^2$ in the second term on the right hand 
side of \Eq{eq:v4}, accordingly the coefficient $\chi_4$ 
is found independent of initial eccentricities by construction. Similar strategy can be 
applied to other higher order harmonic flow as well.
For $V_5$, $V_6$ and $V_7$, the corresponding expansion leads to 
\begin{subequations}
\label{eq:vn}
\begin{align}
V_5&=V_5^L + \chi_5 V_2V_3\,,\\
V_6&=V_6^L + \chi_{63}V_3^2 + \chi_{62} V_2^3\,,\\
V_7&=V_7^L + \chi_7 V_2^2V_3\,.
\end{align}
\end{subequations}
The nonlinear terms in the right-hand side are the lowest-order terms involving $V_2$ and $V_3$
which are compatible with rotational symmetry. 
For $V_6$, there exists a non-negligible component from 
cubic order hydro response to $\E_2^3$, which has 
already been noticed through the observed 
event-plane correlation between $V_6$ and $V_2$~\cite{Aad:2014fla},
thus one must expand $V_6$ to cubic order, with an extra
cubic order coefficient $\chi_{62}$. For $V_7$, there is no
contributions of quadratic order, thus the coefficient $\chi_7$ 
is defined regarding the cubic order hydro response to $\E_2^2\E_3$. 

In the hydro response formalism,
$\chi_n$'s are interpreted as ratios between 
nonlinear and linear flow response coefficients, which 
are independent of the initial
density profile for a given centrality class. 
Each $\chi_n$ can be readily evaluated in a single-shot hydrodynamic simulation~\cite{Teaney:2012ke}
by choosing an initial density profile such that only the term involving $\chi_n$ 
is nonvanishing in the expansion of  \Eqs{eq:v4} and (\ref{eq:vn}).
If one analyses a set of events in a centrality class, where the flow fluctuates event to event, 
as in actual heavy-ion experiments and AMPT simulations~\cite{Lin:2004en}, 
$\chi_n$'s can be isolated 
using \Eqs{eq:v4} and (\ref{eq:vn}) under the assumption that the 
terms in the right-hand side are mutually uncorrelated~\cite{Yan:2015jma}:
\begin{subequations}
\label{nonlinear}
\begin{align}
\chi_4&=\frac{\bra V_4 (V_2^*)^2\ket}{\bra |V_2|^4\ket}
=\frac{v_4\{\Psi_2\}}{\sqrt{\bra |V_2|^4\ket}}\\
\chi_5&=\frac{\bra V_5 V_2^*V_3^*\ket}{\bra |V_2|^2 |V_3|^2\ket}
=\frac{v_5\{\Psi_{23}\}}{\sqrt{\bra |V_2|^2 |V_3|^2\ket}}\\
\chi_{62}&=\frac{\bra V_6 (V_2^*)^3\ket}{\bra |V_2|^6\ket}
=\frac{v_6\{\Psi_2\}}{\sqrt{\bra |V_2|^6\ket}}\,,\qquad
\chi_{63}=\frac{\bra V_6 (V_3^*)^2\ket}{\bra |V_3|^4\ket}
=\frac{v_6\{\Psi_3\}}{\sqrt{\bra |V_3|^4\ket}}\\
\chi_7&=\frac{\bra V_7(V_2^*)^2V_3^*\ket}{\bra |V_2|^4 |V_3|^2\ket}
=\frac{v_7\{\Psi_{23}\}}{\sqrt{\bra |V_2|^4 |V_3|^2\ket}}.
\end{align}
\end{subequations}    
The expressions on the right hand side of \Eqs{nonlinear} 
involve the higher order harmonic flow measured in the plane
of lower order harmonics. For example, $V_4$ can be measured in experiments
in its own event-plane $\Psi_4$ which is defined in \Eq{Vndef}, as well as
the event-plane $\Psi_2$ which is determined by $V_2$. More explicitly,
$V_4$ measured with respect to $\Psi_2$ is
\be
\label{eq:v4psi2}
v_4\{\Psi_2\} \equiv \frac{Re\bra V_4 (V_2^*)^2\ket}{\sqrt{\bra |V_2|^4\ket}}
= \bra \cos 4(\Psi_4-\Psi_2)\ket_w \times v_4\{\Psi_4\}\,.
\ee
It is worth mentioning that measuring higher order flow harmonics in
the event plane of $V_2$ and/or $V_3$ is 
equivalent to the corresponding measurement
of event plane correlations~\cite{Aad:2014fla},
as demonstrated by the second identity in \Eq{eq:v4psi2}. 
The denominators in \Eqs{nonlinear} involve various moments of 
the distributions of $V_2$ and $V_3$. There is no direct measurement
of flow moments up to date in experiments, though it can be done in
a generalized scalar-product method, with sufficient rapidity gap~\cite{Bhalerao:2014xra}.
In this work, we extract flow moments from 
flow cumulants~\cite{Borghini:2001vi} which are measured. For instance,
the fourth order moment of $V_2$ is related to $v_2\{2\}$ and $v_2\{4\}$
by
\be
\bra |V_2|^4 \ket = 2v_2\{2\}^4-v_2\{4\}^4
\ee

\section{Results and discussions}
\label{sec:res}

\begin{figure}
\includegraphics[width=1.00\textwidth] {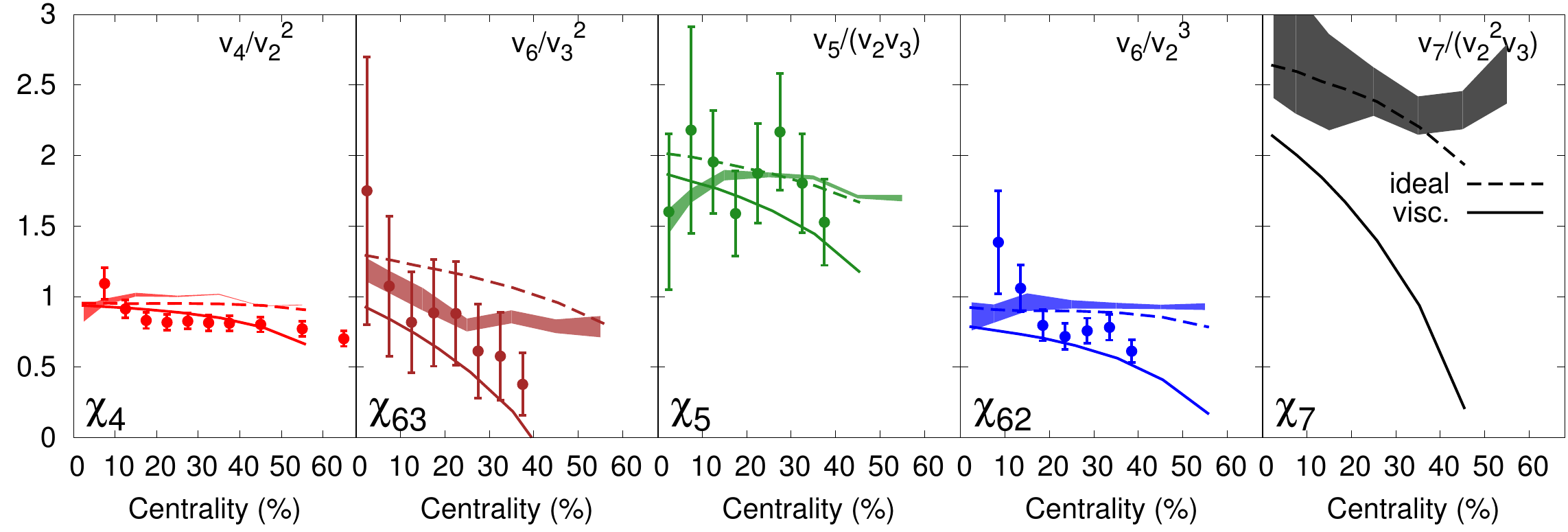}
\caption{
\label{fig:multi}
(Color online) Nonlinear response coefficients $\chi_n$ 
as a function of centrality percentile for $\sqrt{s_{NN}}=2.76$ TeV PbPb
at the LHC. Symbols with errors are extracted results from CMS and ATLAS collaborations. 
Lines are from single-shot hydro simulations with $\eta/s=1/4\pi$
(solid line) or $\eta/s=0$ (dashed line). Shaded bands are results from AMPT
simulations with parton ellastic cross-section $\sigma=1.5$ mb.
}
\end{figure}

The CMS collaboration has measured $v_4\{\Psi_2\}$, $v_6\{\Psi_2\}$ and cumulants of $V_2$
distributions~\cite{Chatrchyan:2013kba}\footnote{
Cumulants of $V_2$ distributions from the CMS collaboration
have so far been published up to $v_2\{4\}$. In this work
we approximately take $v_2\{8\}\approx v_2\{6\}\approx v_2\{4\}$, as being implied from
the measurements by the ATLAS collaboration~\cite{Aad:2013xma}. 
}, 
therefore we are able to assess $\chi_4$ and $\chi_{62}$ according 
to \Eqs{nonlinear}. To evaluate $\chi_5$ and $\chi_{63}$, we estimate $v_5\{\Psi_{23}\}$ and
$v_6\{\Psi_3\}$ from 
the event-plane correlations 
$\bra\cos(5\Psi_5-2\Psi_2-3\Psi_3)\ket_w$ and $\bra\cos6(\Psi_6-\Psi_3)\ket_w$
measured by the ATLAS collaboration~\cite{Aad:2014fla}, in addition to the cumulants of $V_3$ distributions
from the CMS collaboration. $\chi_n$'s from the $\sqrt{s_{NN}}=2.76$ TeV 
PbPb at the LHC are shown as symbols in \Fig{fig:multi}. 
To make comparisons, we calculate these nonlinear response coefficients 
from single-shot hydro simulations~\cite{Teaney:2012ke} as 
well as AMPT~\cite{Lin:2004en}. There is no event-by-event 
fluctuations implemented in our hydro simulations, 
where the initial condition is taken by perturbing a 
smooth and azimuthally symmetric Gaussian density profile with specific initial
eccentricities. 
The normalization of the Gaussian profile is
adjusted to fit the values of $dN_{\small{ch}}/dy$ of LHC PbPb in a given centrality class.  
Results from ideal and viscous (with $\eta/s=1/4\pi$) hydro simulations are depicted as 
dashed and solid lines respectively in \Fig{fig:multi}, which present an overall 
agreement comparing with the experimental data.
AMPT simulations contain non-trivial event-by-event fluctuations at the nucleonic
and partonic levels, and the parton ellastic cross-section is taken to be
$\sigma=1.5$ mb.
It is worth mentioning that ideal hydrodynamics predicts quantitative relations 
among these nonlinear response coefficients due to Cooper-Fyer freeze-out~\cite{Borghini:2005kd}: 
$\chi_4\sim\chi_{63}\sim\frac{1}{2}\chi_5$,
and $\chi_{62}\sim \frac{1}{3}\chi_7$, which are consistent with
the experimental data in \Fig{fig:multi} as well as the results obtained
from AMPT. 
\begin{figure}
\begin{center}
\includegraphics[width=0.40\textwidth] {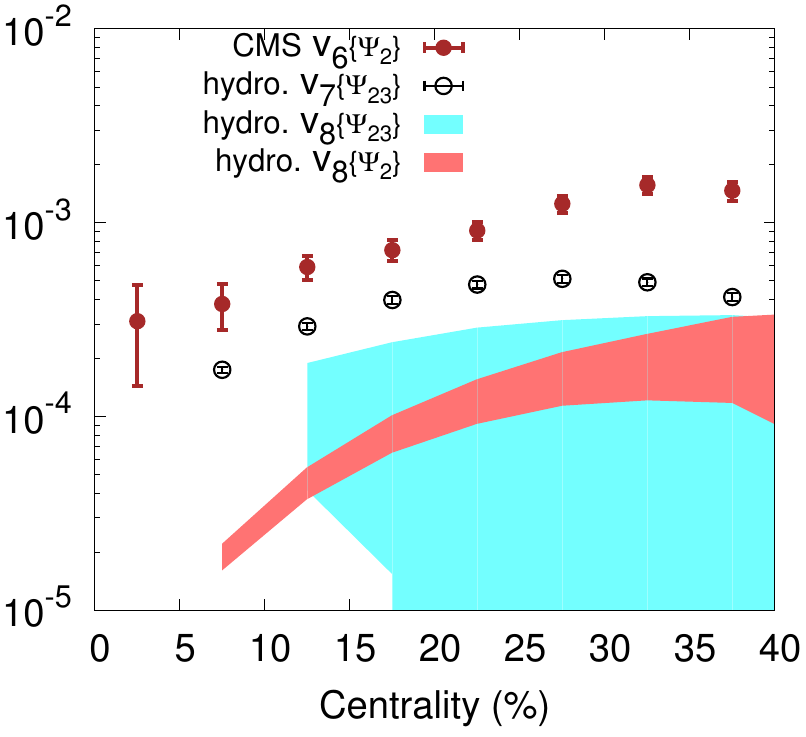}
\caption{
\label{fig:chi8}
(Color online) Hydro predictions for $v_7$ and $v_8$ measured 
in the event-plane constructed by $v_2$ and $v_3$. Boundaries of 
the colored bands are determined by ideal and viscous ($\eta/s=1/4\pi$)
hydro calculations respectively. 
}
\end{center}
\end{figure}

Although there is no experimental data 
available so far for the extraction of 
nonlinear response coefficients $\chi_n$ for the flow harmonics of order
$n>6$, we make predictions in our model simulations for $V_7$ 
and $V_8$.
For $V_8$, there exist a cubic order term and a 
quartic order term allowed by rotational symmetry,
\be
V_8=V_8^L+\chi_{8(23)}\E_2\E_3^2 + \chi_{8(2)}\E_2^4\,,
\ee 
which correspond in experiments to the measurements of $V_8$
in the event-plane constructed by $V_2$ and $V_3$,
and event-plane of $V_2$ respectively,
\be
\label{eq:v8}
\chi_{8(23)}=\frac{v_8\{\Psi_{23}\}}{\sqrt{\bra |V_2|^2\ket\bra |V_3|^4\ket}}\,,\qquad
\chi_{8(2)}=\frac{v_8\{\Psi_{2}\}}{\sqrt{\bra |V_2|^8\ket}}
\ee
$\chi_7$ from our model simulations are presented in \Fig{fig:multi}.
In \Fig{fig:chi8}, $V_7$ and $V_8$ measured in the event-plane consturcted 
by $V_2$ and $V_3$ are predicted with single-shot hydro simulations
according to Eqs.~(\ref{nonlinear}d) and (\ref{eq:v8}).


\section{Conclusions}
\label{sec:con}

Under fairly general assumptions, we have proposed a new set of
nonlinear response coefficients $\chi_n$ based on the measurements of 
higher order harmonic flow with respect to the event-plane 
constructed by $v_2$ and $v_3$. These coefficients are independent
of the detailed information of initial state by construction.
Model simulations with single-shot
hydrodynamics and AMPT give rise to predictions in good agreement with 
experimental data. We noticed that the relative ratios among these
coefficients are consistent with an ideal hydro expectation based on
the analysis of freeze-out. Nonlinear response coefficients associated
with $v_7$ and $v_8$ are calculated as well in our theoretical models as
predictions.  

\section*{Acknowledgements}
LY is funded by the European Research Council under
the Advanced Investigator Grant ERC-AD-267258.






\bibliographystyle{iopart-num}
\bibliography{qm15_yan}







\end{document}